\documentclass{article}

\usepackage{amsmath,cite,bm}

\newcommand{\rd}{\mathrm{d}}
\begin{document}
\begin{titlepage}
\begin{center}
{\large \textbf{Autonomous models solvable through the full interval method }}

\vspace{2\baselineskip}
{\sffamily Mohammad~Khorrami~\footnote{e-mail: mamwad@mailaps.org},
Amir~Aghamohammadi~\footnote{e-mail: mohamadi@alzahra.ac.ir}.}

\vspace{2\baselineskip}
{\it Department of Physics, Alzahra University, Tehran 19938-93973, Iran}
\end{center}
\vspace{2\baselineskip}
\begin{abstract}
\noindent The most general exclusion single species
one dimensional reaction-diffusion models with nearest-neighbor
interactions which are both autonomous and  can be solved exactly
through full interval method are introduced. Using
a generating function method, the general solution for, $F_n$,
the probability that $n$ consecutive sites be full, is obtained.
Some other correlation functions of number operators at
nonadjacent sites are also explicitly obtained. It is shown that
for a special choice of initial conditions some correlation functions
of number operators called full intervals remain uncorrelated.
\end{abstract}
\end{titlepage}
\newpage
\section{Introduction}
Most of the analytical studies on non-equilibrium statistical models
belong to the low-dimensional (specially one dimensional) models \cite{ScR,HHL1,MHP2,ADHR,KPWH,HS1,PCG,HOS1,HOS2,AL,AKK,AKK2,AM1}.
Analyzing one dimensional models, which are usually easier
to investigate, helps us gaining knowledge on systems
far from equilibrium. One of the techniques used to obtain
exact results is the empty interval method (EIM), or its equivalent
the full interval method (FIM). Among other things, it has been
used to analyze the one dimensional dynamics of diffusion-limited
coalescence \cite{BDb,BDb1,BDb2,BDb3}. In these, one dimensional
diffusion-limited processes have been studied using EIM. There, some of
the reaction rates have been taken infinite, and the models
have been worked out on continuum. For the cases of finite
reaction-rates, some approximate solutions have been obtained.
Using this method $E_n$ (the probability that $n$ consecutive
sites be empty) has been calculated. (Alternatively, $F_n$ is
the probability that $n$ consecutive sites be full.)
This method has been used to study a reaction-diffusion
process with three-site interactions \cite{HH}. EIM has been
also generalized to study the kinetics of the q-state
one dimensional Potts model in the zero-temperature limit \cite{Mb}.
In \cite{AKA}, all the one dimensional reaction-diffusion
models with nearest neighbor interactions which can be
exactly solved by EIM have been found and studied. Conditions
have been obtained for the systems with finite reaction
rates to be solvable via EIM, and then the equations
of EIM have been solved. There solvability means that
the evolution equation for $E_n$ be closed. It turned out
there, that certain relations between the reaction rates are needed,
so that the system is solvable via EIM. When these conditions
between reaction rates are met, the time derivative of
$E_n$'s would be linear combinations of $E_n$'s. It was shown
that if certain reactions are absent, namely reactions that
produce particles in two adjacent empty sites, the coefficients
of the empty intervals in the evolution equation of
the empty intervals are $n$-independent, so that the evolution
equation can be easily solved. The criteria for solvability,
and the solution of the empty-interval equation were
generalized to cases of multi-species systems and multi-site
interactions in \cite{KAA,AAK,AK}. In \cite{AK2}, models were
studied which were solvable through EIM, but did include
interaction which produce particles in two adjacent empty sites.
There these models were investigated in continuum, although
some terms in the evolution equation were missed, as will be
discussed in the present paper. In \cite{MMPB}, conventional
EIM has been extended to a more generalized form. Using
this extended version, a model has been studied which can not
be solved by conventional EIM. Recently,
the coagulation-diffusion process on a one dimensional
chain has been studied using the empty-interval method \cite{DFBHR}.
There the behavior of the time-dependent double-empty-interval
probability has been studied. In \cite{DFH}, the exact two-time
correlation and response functions for a one dimensional
coagulation-diffusion process has been studied using EIM.

In \cite{Sch}, a ten-parameter family of reaction-diffusion
processes was introduced for which the evolution equation
of $n-$point functions contains only $n-$ or less-point
functions. We call such systems autonomous. The expectation value
of the particle-number in each site has been obtained exactly
for these models. In order to be an autonomous model there should be
some constraints on the reaction rates.

Here we study the most general exclusion single species
one dimensional reaction-diffusion model with
nearest-neighbor interactions, which can be solved
exactly through the full interval method, and is autonomous.
The reaction rates corresponding to these two models
apart from corrections to \cite{AK2}, here lattice models
are studied, while in \cite{AK2} such models on continuum
were investigated. The change from the empty interval to
the full interval is, of course, not important; as a simple
interchange of particles and holes would do that. The scheme
of the paper is as follows. In Section 2, the most general
exclusion single species one dimensional reaction-diffusion
models with nearest-neighbor interactions are introduced,
which are autonomous and can be solved exactly through FIM.
In Section 3, the evolution equation for the full interval
probabilities, $F_n$'s are obtained. In section 4,
the steady state solutions for these probabilities are
obtained, and then using a generating function method,
the general solution for $F_n$'s is calculated.
Correlation functions of number operators at nonadjacent sites
are obtained in section 5. Finally, in section 6 some
correlation functions of number operators are explicitly
calculated for some special choice of initial conditions.
These are probabilities of some disjoint parts of the lattice
be full.
\section{Full interval and autonomy}
Consider a one dimensional lattice, any site of which is either
occupied by a single particle or empty, and assume that the reactions, 
as well as the state of the system, are translationally invariant. 
Implicit in this, is that the lattice has no boundaries. But 
the lattice can still be finite, if it is circular. 
Defining $F_n$ as the probability that $n$ consecutive sites be full
\begin{equation}\label{03.1}
F_n:=P(\overbrace{\bullet\bullet\cdots\bullet }^n),
\end{equation}
where an empty (occupied) site is denoted by by $\circ$ ($\bullet$),
it is found (for example similar to \cite{AK2}), that
the most general single species nearest-neighbor interactions
for which the evolution equations governing $F_n$'s are closed are
\begin{align}\label{03.2}
\circ\bullet&\to\begin{cases}\circ\circ,& q_1\\
                              \bullet\circ,& r_1
\end{cases},\nonumber\\
\bullet\circ&\to\begin{cases}\circ\circ,& q_2\\
                           \circ\bullet,& r_2
\end{cases},\nonumber\\
\circ\circ&\to\begin{cases}\bullet\circ,& r_1\\
                    \circ\bullet,& r_2
\end{cases},\nonumber\\
\bullet\bullet&\to\begin{cases}\bullet\circ,&w_1\\
                              \circ\bullet,&w_2\\
                               \circ\circ,&w
\end{cases}.
\end{align}
where $r_1, r_2, q_1, q_2, w, w_1$ and $w_2$ are reaction rates.
Then the Hamiltonian for a two site interaction for models solvable through FIM is
\begin{equation}\label{03.3}
H=\begin{pmatrix}
-r_1-r_2&q_1&q_2&w\\
r_2&-q_1-r_1&r_2&w_2\\
r_1&r_1&-q_1-r_1&w_1\\
0&0&0&-w-w_1-w_2
\end{pmatrix}.
\end{equation}
The autonomy criteria leads to two more constraints (\cite{Sch}, for example)
\begin{align}\label{03.4}
r_1+q_1&=w_1+w,\nonumber\\
r_2+q_2&=w_2+w.
\end{align}
So an autonomous model solvable through FIM has five free parameters.
\section{The full interval evolution}
The full interval equation is
\begin{align}\label{03.5}
\frac{\rd F_n}{\rd t}&=(r_1+r_2)\,(F_{n-1}+F_{n+1}-2\,F_n)-(q_1+q_2)\,(F_n-F_{n+1})\nonumber\\
&\quad-(n-1)\,(w_1+w_2+w)\,F_n-(w_1+w_2+2\,w)\,F_{n+1},
\end{align}
where
\begin{equation}\label{03.6}
F_0:=1.
\end{equation}
The difference of this with what obtained in \cite{AK2}, apart from
the obvious interchange of particles and vacancies, is that
the last term had been missed in \cite{AK2}.
Using (\ref{03.4}), it is seen that the coefficient of $F_{n+1}$
in the right-hand side of (\ref{03.5}) vanishes. So,
\begin{equation}\label{03.7}
\frac{\rd F_n}{\rd t}=(r_1+r_2)\,F_{n-1}-[2\,(r_1+r_2)+q_1+q_2+(n-1)\,(w_1+w_2+w)]\,F_n.
\end{equation}
Rescaling the time by $(w_1+w_2+w)$:
\begin{equation}\label{03.8}
\tilde t:=(w_1+w_2+w)\,t,
\end{equation}
the evolution equation becomes
\begin{equation}\label{03.9}
\frac{\rd F_n}{\rd\tilde t}=b\,F_{n-1}-(a+n-1)\,F_n,
\end{equation}
where
\begin{align}\label{03.10}
a&:=\frac{2\,(r_1+r_2)+q_1+q_2}{w+w_1+w_2},\nonumber\\
b&:=\frac{r_1+r_2}{w+w_1+w_2}.
\end{align}
Equation (\ref{03.4}), and the fact that the rates are nonegative
guarantee that $(w_1+w_2+w)$ is positive, unless all of
the rates are zero. These also show that
\begin{align}\label{03.11}
a&\geq 1+b,\\ \label{03.12}
b&\geq 0.
\end{align}
From now on, the rescaled time $\tilde t$ is denoted by $t$,
so that the evolution equation is written as
\begin{equation}\label{03.13}
\dot F_n=b\,F_{n-1}-(a+n-1)\,F_n,
\end{equation}
where dot means differentiation with respect to the rescaled time.
\section{The general solution for the full interval}
Let's first consider the large time values for $F_n$ or the steady state solution. Denoting the time independent equation for the full interval by
$F_n^\mathrm{st}$, one has
\begin{equation}\label{03.14}
F_n^\mathrm{st}=\frac{b}{a+n-1}\,F_{n-1}^\mathrm{st},
\end{equation}
which combined with (\ref{03.6}) results in
\begin{equation}\label{03.15}
F_n^\mathrm{st}=\frac{b^n\,\Gamma(a)}{\Gamma(a+n)}.
\end{equation}

The general solution to (\ref{03.13}) is of the form
\begin{equation}\label{03.16}
F_n(t)=F_n^\mathrm{st}+\sum_{m=1}^n c_{n\,m}\,\exp[-(a+m-1)\,t],
\end{equation}
where $c_{n\,m}$'s are constants. Putting (\ref{03.16}) in
(\ref{03.13}), one arrives at
\begin{equation}\label{03.17}
(n-m)\,c_{n\,m}=b\,c_{n-1\,m},\qquad m<n.
\end{equation}
This allows one to determine $c_{n\,m}$'s in terms of
$c_{m\,m}$'s, which are denoted by $d_m$:
\begin{align}\label{03.18}
c_{n\,m}&=\frac{b^{n-m}}{(n-m)!}\,c_{m\,m},\nonumber\\
&=:\frac{b^{n-m}}{(n-m)!}\,d_m.
\end{align}
So,
\begin{equation}\label{03.19}
F_n(t)=F_n^\mathrm{st}+\sum_{m=1}^n \frac{b^{n-m}}{(n-m)!}\,d_m\,\exp[-(a+m-1)\,t],
\end{equation}
showing that the (largest) relaxation time $\tau$ is obtained from
\begin{equation}\label{03.20}
\tau=\frac{1}{a}.
\end{equation}
The constants $d_m$ are to be obtained from the initial conditions.
One way is to define the generating functions $G$ and $G^\mathrm{st}$ through
\begin{align}\label{03.21}
G&:=\sum_{n=0}^\infty F_n(0)\,x^n,\nonumber\\
G^\mathrm{st}&:=\sum_{n=0}^\infty F_n^\mathrm{st}\,x^n.
\end{align}
Using these and (\ref{03.19}),
\begin{align}\label{03.22}
G(x)&=G^\mathrm{st}(x)+\sum_{n=0}^\infty\sum_{m=1}^n \frac{b^{n-m}}{(n-m)!}\,d_m\,x^n,\nonumber\\
&=G^\mathrm{st}(x)+\sum_{m=1}^\infty\sum_{n=m}^\infty \frac{b^{n-m}}{(n-m)!}\,d_m\,x^n,\nonumber\\
&=G^\mathrm{st}(x)+\sum_{m=1}^\infty\left(\frac{\partial}{\partial b}\right)^m\sum_{n=0}^\infty \frac{b^n}{n!}\,d_m\,x^n,\nonumber\\
&=G^\mathrm{st}(x)+\sum_{m=1}^\infty d_m\,\left(\frac{\partial}{\partial b}\right)^m\exp(b\,x),\nonumber\\
&=G^\mathrm{st}(x)+\exp(b\,x)\,\sum_{m=1}^\infty d_m\,x^m.
\end{align}
So,
\begin{equation}\label{03.23}
\sum_{m=1}^\infty d_m\,x^m=\exp(-b\,x)\,[G(x)-G^\mathrm{st}(x)],
\end{equation}
which results in
\begin{equation}\label{03.24}
d_m=\sum_{k=0}^{m-1}\frac{(-b)^k}{k!}\,\left[F_{m-k}(0)-\frac{b^{m-k}\,\Gamma(a)}{\Gamma(a+m-k)}\right].
\end{equation}
For example, the occupation probability of any site is
\begin{equation}\label{03.25}
F_1(t)=\frac{b}{a}+\left[F_1(0)-\frac{b}{a}\right]\,\exp(-a\,t).
\end{equation}
\section{Correlation functions of number operators at nonadjacent site}
Denoting the number operator in the site $i$ by $n_i$, it is seen that
\begin{equation}\label{03.26}
\langle\dot n_i\rangle=b-a\,\langle n_i\rangle.
\end{equation}
This is in fact the same as the evolution equation for $F_1$, as it should be.
The solution to (\ref{03.26}) is
\begin{equation}\label{03.27}
\langle n_i\rangle(t)=\langle n_i\rangle(0)\,\exp(-a\,t)+\frac{b}{a}\,[1-\exp(-a\,t)].
\end{equation}
Defining the correlation $C_{i_1\cdots i_k}$ as
\begin{equation}\label{03.28}
C_{i_0\cdots i_k}:=\langle n_{i_0}\cdots n_{i_k}\rangle,
\end{equation}
where no two indices are adjacent, it is seen that
\begin{equation}\label{03.29}
\dot C_{i_0\cdots i_k}=b\sum_{j=0}^k C_{i_0\cdots\widehat{i_j}\cdots i_k}
-(k+1)\,a\,C_{i_0\cdots i_k},
\end{equation}
where $\hat i$ means that the index $i$ has been omitted.

A simple change of variable makes the above equations simpler.
Defining
\begin{equation}\label{03.30}
\tilde C_{i_0\cdots i_k}:=\left\langle\left(n_{i_0}-\frac{b}{a}\right)\cdots
\left(n_{i_k}-\frac{b}{a}\right)\right\rangle,
\end{equation}
(again for the case no two indices are adjacent) one arrives at
\begin{equation}\label{03.31}
\dot{\tilde C}_{i_0\cdots i_k}=-(k+1)\,a\,\tilde C_{i_0\cdots i_k}.
\end{equation}
The so called connected correlations $\tilde C^\mathrm{c}$ are
defined inductively through
\begin{align}\label{03.32}
\tilde C_i&=:\tilde C^\mathrm{c}_i,\\ \label{03.33}
\tilde C_{i_0\cdots i_k}&=:\sum_{\mathcal{P}}\tilde C^\mathrm{c}_{\mathcal{P}_1(i_0\cdots i_k)}\cdots
\tilde C^\mathrm{c}_{\mathcal{P}_\alpha(i_0\cdots i_k)},
\end{align}
where the summation runs over all partitions $\mathcal{P}$
of $(i_0\cdots i_k)$. Such a partition divides the indices
$(i_0\cdots i_k)$ into $\alpha$ parts, where the $\beta$'th part
is denoted by $\mathcal{P}_\beta(i_0\cdots i_k)$.
A simple induction shows that
\begin{equation}\label{03.34}
\dot{\tilde C}^\mathrm{c}_{i_0\cdots i_k}=-(k+1)\,a\,\tilde C^\mathrm{c}_{i_0\cdots i_k}.
\end{equation}
Another induction shows that for $k>0$, adding a constant to any of
$n_{i_j}$'s does not change the connected correlations. The reason is
that, denoting the correlations and connected correlations corresponding
to $(n_{i_k}+\Delta)$ by a superscript $\Delta$, one has
\begin{equation}\label{03.35}
\tilde C^{\Delta\,\mathrm{c}}_{i_0\cdots i_k}=\tilde C^\Delta_{i_0\cdots i_k}-
\tilde C^\Delta_{i_0\cdots i_{k-1}}\,\tilde C^\Delta_{i_k}+R,
\end{equation}
where $R$ contains terms in which the index $i_k$ enters
connected correlations of $(m+1)$~operators, where $m$ is positive
but less than $k$. Assuming that the independence of the connected
correlations of $(m+1)$~operators is true for $m$ positive and
less than $k$, it is seen that $R$ does not depend on $\Delta$.
The sum of the first two terms of the right hand side is also
obviously independent of $\Delta$. So the left hand side is
independent of $\Delta$ as well. To complete the induction,
one should prove that the connected correlation of two number
operators does not change upon adding a constant term to
each of them. This is obvious since in that case, $R$ in the
right hand side of (\ref{03.35}) is zero.

In particular, one arrives at
\begin{equation}\label{03.36}
C^\mathrm{c}_{i_0\cdots i_k}=\tilde C^\mathrm{c}_{i_0\cdots i_k},\qquad k>0,
\end{equation}
showing that
\begin{equation}\label{03.37}
C^\mathrm{c}_{i_0\cdots i_k}(t)=C^\mathrm{c}_{i_0\cdots i_k}(0)\,\exp[-(k+1)\,a\,t],\qquad k>0.
\end{equation}
The simplest example of this, is the connected two point function:
\begin{equation}\label{03.38}
(\langle n_i\,n_j\rangle-\langle n_i\rangle\,\langle n_j\rangle)(t)=
(\langle n_i\,n_j\rangle-\langle n_i\rangle\,\langle n_j\rangle)(0)\,\exp(-2\,a\,t),\quad|j-i|>1.
\end{equation}

To consider correlations with possibly adjacent sites, let us
first obtain the evolution equation for the full interval, without
the assumption of translational invariance. Defining $F_{i\,j}$
as the probability that the sites beginning from $i$ ending in
$j$ are full, one arrives at
\begin{equation}\label{03.39}
\dot F_{i\,j}=b_1\,F_{i\,j-1}+b_2\,F_{i+1\,j}-(a+j-i)\,F_{i\,j},
\end{equation}
where
\begin{align}\label{03.40}
b_1&:=\frac{r_1}{w+w_1+w_2},\nonumber\\
b_2&:=\frac{r_2}{w+w_1+w_2}.
\end{align}
Then, reassuming translational invariance define $D_{\bm{\ell}}$ as
\begin{equation}\label{03.41}
D_{\bm{\ell}}:=\left\langle\prod_{j=0}^k\left(\prod_{m=1}^{\ell_{2\,j}}
n_{\ell_0+\cdots+\ell_{2\,j-1}+m}\right)\right\rangle,
\end{equation}
where
\begin{equation}\label{03.42}
\bm{\ell}:=(\ell_0,\dots,\ell_{2\,k}),
\end{equation}
and $\ell_j$'s are positive integers.
This is the probability that $\ell_0$ consecutive sites beginning
from $1$ be full, $\ell_2$ consecutive sites beginning from $(\ell_0+\ell_1+1)$
be full, and so on. In the special case $k=0$, this correlation is the same as
the full interval:
\begin{equation}\label{03.43}
D_\ell=F_\ell.
\end{equation}
In the special case that all $\ell_j$'s with even $j$ are one, $D$
is reduced to the $(k+1)$-point function with nonadjacent sites:
\begin{equation}\label{03.44}
D_{1,\ell_1\,1,\dots,\ell_{2\,k-1},1}=C_{i_0\cdots i_k},
\end{equation}
where
\begin{equation}\label{03.45}
i_{m+1}-i_m=\ell_{2\,m+1}+1.
\end{equation}
Using (\ref{03.39}) and (\ref{03.41}), it is then seen that
\begin{align}\label{03.46}
\dot D_{\bm{\ell}}&=\sum_{m=0}^k[b_2\,D_{\bm{\ell}-\bm{e}_{2\,m}+\bm{e}_{2\,m-1}}
+b_1\,D_{\bm{\ell}-\bm{e}_{2\,m}+\bm{e}_{2\,m+1}}-(a+\ell_{2\,m}-1)\,D_{\bm{\ell}}],
\end{align}
where $\bm{e}_m$ is a $(2\,k+1)$-tuple, the only nonzero component
of which is the $m$-th component being equal to one, and
$\bm{e}_{-1}$ and $\bm{e}_{2\,k+1}$ are zero.

A special case is the two-point function. For nonadjacent sites,
one use
\begin{equation}\label{03.47}
\dot C_{i\,j}=b\,(C_i+C_j)-2\,a\,C_{i\,j},
\end{equation}
the solution to which is
\begin{align}\label{03.48}
C_{i\,j}(t)&=\langle n_i\,n_j\rangle(0)\,\exp(-2\,a\,t)+\frac{b^2}{a^2}\,[1-\exp(-2\,a\,t)]\nonumber\\
&\quad+\frac{b}{a}\,\left[\langle n_i\rangle(0)+\langle n_j\rangle(0)-\frac{2\,b}{a}\right]\,
[\exp(-a\,t)-\exp(-2\,a\,t)],
\end{align}
where (\ref{03.27}) has been used, and the fact that
\begin{equation}\label{03.49}
C_i(t)=\langle n_i\rangle(t).
\end{equation}
Using (\ref{03.39}), one has
\begin{equation}\label{03.50}
\dot F_{i\,j}=b_1\,C_i+b_2\,C_j-(a+1)\,F_{i\,j},\qquad j-i=1,
\end{equation}
the solution to which is
\begin{align}\label{03.51}
F_{i\,j}(t)&=\langle n_i\,n_j\rangle(0)\,\exp[-(a+1)\,t]+\frac{b^2}{a\,(a+1)}\,\{1-\exp[-(a+1)\,t]\}
\nonumber\\&\quad+\left[b_1\,\langle n_i\rangle(0)+b_2\,\langle n_j\rangle(0)-\frac{b^2}{a}\right]
\nonumber\\&\qquad\times\{\exp(-a\,t)-\exp[-(a+1)\,t]\},\qquad j-i=1.
\end{align}
One thus arrives for an expression for the two point functions:
\begin{equation}\label{03.52}
\langle n_i\,n_j\rangle(t)=\begin{cases}
C_{i\,j}(t),&j-i>1\\
F_{i\,j}(t),&j-i=1\\
\langle n_i\rangle(t),&j-i=0
\end{cases}.
\end{equation}
\section{Uncorrelated full intervals}
A special case of the initial conditions is when
\begin{equation}\label{03.53}
D_{\bm{\ell}}(0)=\prod_{j=0}^k\left\langle\prod_{m=1}^{\ell_{2\,j}}
n_{\ell_0+\cdots+\ell_{2\,j-1}+m}\right\rangle(0),\qquad\forall\;\bm{\ell},
\end{equation}
which can be written as
\begin{equation}\label{03.54}
D_{\bm{\ell}}(0)=\prod_{j=0}^k F_{\ell_{2\,j}}(0),\qquad\forall\;\bm{\ell}.
\end{equation}
Using (\ref{03.13}), it is seen that the ansatz
\begin{equation}\label{03.55}
D_{\bm{\ell}}(t)=\prod_{j=0}^k F_{\ell_{2\,j}}(t),
\end{equation}
does satisfy (\ref{03.46}). This does not mean that the system is
completely uncorrelated, as the ansatz
\begin{equation}\label{03.56}
F_n(t)=(F_1)^n(t),
\end{equation}
does not satisfy (\ref{03.13}). But it means that the full intervals
are uncorrelated to each other. Among other things, it does mean
that the correlators $C$ satisfy
\begin{equation}\label{03.57}
C_{i_0\cdots i_k}=(F_1)^{k+1},
\end{equation}
if this holds initially.
\\[\baselineskip]
\textbf{Acknowledgement}:  This work was supported by
the research council of the Alzahra University.

\end{document}